\documentclass[aps,prl,twocolumn,groupedaddress,floatfix,showpacs]{revtex4}
\usepackage{graphicx}
\usepackage{amsmath}

\begin{document}

\title{Adiabatic pumping through interacting quantum dots}

\author{
Janine Splettstoesser$^{1,2}$, 
Michele Governale$^{1,2}$,
J\"urgen K\"onig$^2$,
and  Rosario Fazio$^1$}

\affiliation{
$^1$NEST-INFM \& Scuola Normale Superiore, I-56126 Pisa, Italy\\
$^2$Institut f\"ur Theoretische Physik III,
Ruhr-Universit\"at Bochum, D-44780 Bochum, Germany}

\date{\today}
\begin{abstract}
We present a general formalism to study adiabatic pumping through interacting 
quantum dots. We derive a formula that relates the pumped charge to the local, 
instantaneous Green's function of the dot. This formula is 
then applied to the infinite-$U$ Anderson model both for weak and 
strong tunnel-coupling strengths.
\end{abstract}

\maketitle

\textit{Introduction}. 
The idea of producing a DC current at zero bias voltage by changing 
some parameters of a conductor periodically in time dates back to the 
work of Thouless \cite{thouless}. 
This method of exploiting the explicit time dependence of the Hamiltonian of the 
system  
is known as \textit{pumping}. If the parameters change slowly as compared to 
all internal time scales
of the system,
pumping is \textit{adiabatic}, and the average transmitted charge does not 
depend on the detailed time dependence of the parameters.  
For non-interacting mesoscopic systems, Brouwer \cite{brouwer}, using 
the concept of emissivity proposed by B\"uttiker \textit{et al.}
\cite{emissivity}, related the charge pumped in 
a period to the derivatives of the instantaneous scattering matrix of the 
conductor with respect to the time-varying parameters. 
In case of noninteracting electrons, a general framework 
for the computation of the pumped charge
has been developed \cite{brouwer,zhou}.
 Pumping through open quantum dots has also been investigated experimentally \cite{marcus}. 

The situation is profoundly different for pumping through interacting 
systems. In fact, there are only few works that address this problem
\cite{pothier,aono,cota} with methods suited to tackle specific systems or regimes. 
As far as pumping through interacting quantum dots is concerned, 
the work by Aono \cite{aono} exploits the zero-temperature mapping of the Kondo problem \cite{raikh} to a noninteracting system and uses the noninteracting formalism. 
On the other hand, Cota \textit{et al.} \cite{cota} study adiabatic pumping 
in a double-dot system in the sequential tunneling limit.

The aim of this Letter is to derive a formula for the charge pumped  
through an interacting quantum dot, which is valid from the high-temperature limit where sequential tunneling dominates down to low temperatures where Kondo correlations are relevant.

\textit{Model and formalism.} We consider a single-level quantum dot 
coupled to two noninteracting leads.
The system is described by the Hamiltonian
\begin{equation}
\label{hamiltonian}
   H = H_\text{leads} + H_\text{dot} + H_\text{tun},
\end{equation}
with $H_{\text{leads}} = \sum_{k,\sigma,\alpha} \epsilon_\alpha(k)   
c_{\sigma k\alpha}^\dagger c_{\sigma k\alpha},$
where $c_{\sigma k\alpha}$ ($c_{\sigma k\alpha}^\dagger $) is the fermionic 
annihilation (creation) operator for an electron with spin 
$\sigma=\uparrow,\downarrow$ and momentum $k$ in lead 
$\alpha = \text{L},\text{R}$. 
The leads are assumed to be in thermal equilibrium with the same chemical 
potential and to have flat bands with constant density of states $\rho_\alpha$.

The quantum dot is described by $H_\text{dot} = [\epsilon + 
\Delta\epsilon(t) ] \sum_\sigma n_\sigma + U n_\uparrow n_\downarrow$ with
$n_\sigma=d^\dagger_\sigma d_\sigma$, where $d_\sigma$ ($d_\sigma^\dagger$) 
is the fermionic annihilation (creation) operator for a dot electron with
spin $\sigma$.
The level position of the dot contains a time-independent part $\epsilon$ and
 a 
time-dependent part, $\Delta\epsilon(t)$.
Coulomb interaction in the dot is described by the on-site energy $U$.
Tunneling is modeled by $H_\text{tun} = \sum_{k,\sigma,\alpha} 
\left[V_{\alpha}(t) c_{\sigma k\alpha}^\dagger d_\sigma+\mbox{H.c.}\right]$
with time-dependent tunnel matrix elements $V_{\alpha}(t)$.
We only allow for the modulus, but not the phase, of $V_{\alpha}(t)$ to vary in
time, since a time-dependent phase would correspond to a bias voltage.

By periodically changing (at least two of) the three quantities $V_{\rm L}(t)$,
$V_{\rm R}(t)$, and $\Delta \epsilon(t)$, a finite charge can be pumped
through the quantum dot.
The charge $Q$ that is pumped after one cycle $\mathcal{T}$ is connected to 
the time-dependent current $J_{\text L}(t)$ flowing through the left barrier 
via the relation $Q=\int_0^{\mathcal{T}} J_{\text{L}}(\tau) d\tau$.
The starting point for our analysis is the exact relation 
that expresses the current in terms of the dot Green's function \cite{jauho}
\begin{eqnarray}
  \nonumber
  J_{\text{L}}(t) &=& - \frac{2e}{\hbar} \, \sum_\sigma \mbox{Im} 
  \left[\frac{\Gamma_{\text{L}}(t,t)}{2} G^<_{\sigma\sigma}(t,t)+
    \int \frac{d\omega}{2 \pi} f(\omega)\right.\\ 
    \label{current}
    & & \left.
    \int dt' e^{-i \omega (t'-t)/\hbar} 
    \Gamma_{\text{L}}(t',t)G^{\text{r}}_{\sigma\sigma}(t,t') \right] \, ,
\end{eqnarray}
with $\Gamma_{\text{L}}(t_1,t)=2 \pi \rho_\text{L} V_{\text{L}}(t) 
V_{\text{L}}^{*}(t_1)$, and $f(\omega)$ is the Fermi function.
The lesser, retarded, and advanced Green's function are defined as usual,
$G^{<}_{\sigma\sigma}(t,t') = i \langle d^\dagger_{\sigma}(t') d_{\sigma}(t)
\rangle$, $G^{\text{r}}_{\sigma\sigma}(t,t')=-i \theta(t-t') \langle
\left\{ d_{\sigma}(t), d^\dagger_{\sigma}(t')\right\} \rangle$, and
$G^{\text{a}}_{\sigma\sigma}(t,t') = \left[ G^{\text{r}}_{\sigma\sigma}(t',t)
\right]^*$.
The Green's functions are diagonal in spin space since tunneling is spin
conserving.
Furthermore, spin degeneracy yields $G_{\uparrow\uparrow}(t,t')=
G_{\downarrow\downarrow}(t,t')\equiv G(t,t')$.
We remark that the Green's functions $G(t,t')$ are defined with a
Hamiltonian that explicitly depends on time.
They are determined by the Dyson equation
\begin{equation}
  \label{dyson}
  \check G (t,t') = \check g(t,t') + \int d t_1 d t_2   
  \, \check G (t,t_1) \check \Sigma (t_1,t_2) \check g(t_2,t') \, ,
\end{equation}
in matrix notation $\check A = \left( \begin{array}{cc} 
A^\text{r} & A^< \\ 0 & A^\text{a} \end{array} \right)$ for the bare,
$\check g$, and full Green's function $\check G$, and the self-energy
$\check \Sigma$.
The latter takes into account the tunnel coupling $\Gamma_\alpha(t)$ to 
the leads, the Coulomb interaction $U$ in the dot,and the time-dependent 
part $\Delta \epsilon(t)$ of the level position.
Note that $\check \Sigma(t_1,t_2) = \check \Sigma(t_1,t_2,\left\{ 
H(\tau)\right\}_{\tau \in [t_1,t_2]} )$ is a functional of the time-dependent
Hamiltonian $H(\tau)$ on the interval $[t_1,t_2]$.

We are interested in the behavior of the self-energy and, thus, the Green's 
function for a slowly varying Hamiltonian $H(\tau)$.
This means that the time scale over which the system parameters are varying is large compared to the lifetime of the system.
To construct the adiabatic expansion of the self-energy we first linearize 
the time dependence of the Hamiltonian, 
$H(\tau) \rightarrow H(t_0)+(\tau - t_0){\dot H}(t_0)$, with respect to 
some fixed time $t_0$,
and expand the self-energy up to linear order in the time derivative, where the time ordering in $H(t_0)$ is still done with respect to time $\tau$.
The relation $\int_{\tau_1}^{\tau_2}\tau \dot{H}(t_0) d \tau  = 
(\tau_1+\tau_2)/2 \int_{\tau_1}^{\tau_2}\dot{H}(t_0) d \tau $, valid for each
segment of time evolution between two 
vertices at times
$\tau_1$ and $\tau_2$ in the self-energy, 
motivates a global replacement of the time variable $\tau$ with the average 
time $(t_1+t_2)/2$ in the self-energy.
This replacement defines an approximation, which we refer to as 
the \textit{average-time approximation}\cite{note_corrections}.
As a result, the dependence of the self-energy on the 
{\em function} $H(\tau)$ over the interval $[t_1,t_2]$ is replaced by 
the dependence on the three times $t_0$, $t_1$, and $t_2$ only,
and we arrive at the adiabatic expansion
$\check\Sigma(t_1,t_2,\left\{ H(\tau)\right\}_{\tau \in [t_1,t_2]} ) 
\rightarrow \check\Sigma_0(t_1,t_2,t_0) +  \check\Sigma_1(t_1,t_2,t_0)$ with
\begin{eqnarray}
  \check \Sigma_0(t_1,t_2,t_0) &=& \check 
  \Sigma(t_1,t_2,\left\{ H(t_0)\right\} ),
\\
\label{sigma1}
  \check \Sigma_1(t_1,t_2,t_0) &=& \left( \frac{t_1+t_2}{2} -t_0 \right)
  \frac{\partial \check \Sigma_0(t_1,t_2,t_0)}{\partial t_0} \, .
\end{eqnarray}
The lowest term in the adiabatic expansion corresponds to
replacing the time-dependent Hamiltonian 
$H(\tau)$ with the constant value $H(t_0)$.
Then, $\check \Sigma_0(t_1,t_2,t_0)$ depends on $t_1$ and $t_2$ only via 
the difference $t_1 - t_2$, and we can introduce the Fourier transform
$\check \Sigma_{0} (\omega,t_0) = \int d(t_1-t_2) 
\exp[i \omega (t_1-t_2)/\hbar]\check \Sigma_{0}(t_1,t_2,t_0)$.

The adiabatic expansion $\check G(t,t') \rightarrow \check G_0(t,t',t_0) 
+ \check G_1(t,t',t_0)$ for the Green's function follows from that for the
self-energy via the Dyson equation Eq.~(\ref{dyson}).
Again, we can introduce Fourier transforms $\check G_{0/1} (\omega,t_0) = 
\int d(t-t') \exp[i \omega (t-t')/\hbar] \check G_{0/1}(t,t',t_0)$.
Since our goal is an adiabatic expansion of the current at time $t$ as given 
in Eq.~(\ref{current}), we choose from now on $t_0=t$.
This results in 
\begin{eqnarray}
  \check G_0 (\omega,t) &=& 
  \left[ \left( \check g(\omega) \right)^{-1} - \check\Sigma_{0}(\omega,t)
  \right]^{-1} \, ,
\\
  \check G_1 (\omega,t) &=& i \hbar
  \frac{\partial \check G_0(\omega,t)}{\partial \omega} 
  \frac{\partial \check\Sigma_{0} (\omega,t)}{\partial t} \check G_0(\omega,t)
\nonumber \\
\label{g1}
  &+& \frac{i\hbar}{2} 
  \check G_0(\omega,t)
  \frac{\partial^2 \check\Sigma_{0} (\omega,t)}{\partial \omega \partial t} 
  \check G_0(\omega,t) \, .
\end{eqnarray}
We specify these matrix equations for the retarded and lesser part and
make use of the equilibrium relations $\Sigma_0^{<}(\omega,t)= 
-2i f(\omega) \mbox{Im}\, \Sigma_{0}^{\text{r}}(\omega,t)$ and 
$G_0^{<}(\omega,t) = -2i f(\omega) \mbox{Im}\, G_{0}^{\text{r}}(\omega,t)$,
where $G_{0}^{\text{r}}(\omega,t) = \left[ \omega - \epsilon 
- \Sigma_0^\text{r}(\omega,t) \right]^{-1}$.
Furthermore, the adiabatic expansion for $\Gamma_{\text{L}}(t',t)$ can be 
constructed as $\Gamma_{\text{L}}(t',t) \rightarrow
\Gamma_{\text{L}}(t) - \frac{t-t'}{2} \dot \Gamma_{\text{L}}(t)$ with
$\Gamma_{\text{L}}(t) \equiv \Gamma_{\text{L}}(t,t)$.
Plugging everything into Eq.~(\ref{current}) we find that the zeroth-order term
of the adiabatic expansion for the current vanishes (as it should since it is
equivalent to time-independent problem at equilibrium).
The first-order correction is given by
\begin{eqnarray}
  \label{adiabaticcurrent}
  J_{\text{L}}(t) &=& -\frac{e}{\pi} \int d\omega \left(- \frac{\partial f}
  {\partial \omega} \right)
  \mbox{Re} \left[ 
  \frac{d}{d t} \left[ \Gamma_{\text{L}}(t) G_{0}^{\text{r}}(\omega,t)
  \right]
  \right. \nonumber \\ && \left. 
  \left( G_{0}^{\text{r}}(\omega,t) \right)^{-1}
  G_0^{\text{a}}(\omega,t)
  \right] . 
\end{eqnarray}
A factor 2 accounts for the spin degeneracy.
Equation~(\ref{adiabaticcurrent}) is the central result of this 
Letter \cite{note3}. 
It generalizes Brouwer's formula \cite{brouwer} to interacting quantum dots. 
We emphasize that this result relies on the average-time approximation
for the self-energy. 
The latter is exact whenever the self-energy contains two vertices (either tunneling 
or interaction) only. This is the case for $U=0$  but also for 
$U\rightarrow \infty$ as long as the self-energy is calculated up to linear 
order in the tunneling coupling $\Gamma$, as well as for arbitrary 
interaction at zero temperature, where the interacting problem can be mapped to a noninteracting one.
We now specialize Eq.~(\ref{adiabaticcurrent}) to the case of weak pumping
due to time-dependent tunneling barriers,
$\Gamma_{\alpha}(t)=\bar{\Gamma}_{\alpha}+\Delta \Gamma_\alpha (t)$, where $\left|\Delta \Gamma_\alpha (t)\right|\ll \bar{\Gamma}_{\alpha}$ at any time $t$.
To the lowest order in $\Delta \Gamma_\alpha (t)$ the charge 
$Q= \int_0^{\cal T} J_\text{L}(\tau) d\tau$ in one period $\cal T$ is
\begin{eqnarray}
  Q = -\frac{e \eta \bar{\Gamma}}{\pi \bar{\Gamma}_{\text{L}} \bar{\Gamma}_{\text{R}}} \int d \omega 
  \left(-\frac{\partial f}{\partial \omega} \right) \frac{\partial\bar{\delta}(\omega)}
{\partial \bar{\Gamma}} \bar{T}(\omega) ,
  \label{weakpumping}
\end{eqnarray}
where 
$\bar{\Gamma}=\bar{\Gamma}_{\text{L}}+\bar{\Gamma}_{\text{R}}$, and 
$\eta=\int_0^\mathcal{T} \dot{\Delta \Gamma_{\text{L}}}(t) 
\Delta \Gamma_{\text{R}} (t) dt$. 
The symbol $\bar{\delta}(\omega)$ denotes the phase of the Green's function 
$\bar{G}_0^{\text{r}}(\omega)=|\bar{G}_0^{\text{r}}(\omega)| 
\exp [i \bar{\delta}(\omega)]$ computed with $\Gamma_{\alpha}(t)$ replaced by $\bar{\Gamma}_{\alpha}$,
and 
$\bar{T}(\omega) = 2\bar{\Gamma}_{\text{L}}\bar{\Gamma}_{\text{R}} / 
\bar{\Gamma} \cdot\text{Im}[\bar{G}_0^{\text{r}}(\omega)]$ 
can be interpreted as the transmission
probability through an interacting quantum dot \cite{transmission}.

\textit{Examples.} 
We now consider only weak pumping with the barriers, and we restrict 
ourselves to the case $\bar{\Gamma}_{\text{L}}=\bar{\Gamma}_{\text{R}}=
\bar{\Gamma}/2$. 
We start by studying the 
noninteracting single-level quantum dot, using Eq.~(\ref{weakpumping}), with 
$\bar{G}_0^{\text{r}}(\omega)=(\omega-\epsilon+\frac{i}{2}\bar{\Gamma})^{-1}$. 
From inspection of Eq.~(\ref{weakpumping}) it is clear that there is no 
pumping in the noninteracting case if the level is resonant ($\epsilon=0$). 
In the high-temperature limit $\beta \bar{\Gamma} \ll 1$ (with $\beta = 
1/k_\text{B}T$), the pumped charge reads $ Q = - \frac{e \eta}{2} f''(\epsilon)$.

We now turn our attention to weak pumping with the barriers in the limit
of large electron-electron interaction, $U\rightarrow \infty$. 
For temperatures larger than the Kondo temperature (defined below), we 
approximate the instantaneous Green's function of the dot within the
equation-of-motion method \cite{eom}. 
Replacing $\Gamma_{\alpha}(t)$ by $\bar{\Gamma}_{\alpha}$  one finds 
$\bar G_0^{\text{r}}(\omega)=\left(1-\bar{\langle n\rangle}\right)\left(\omega-\epsilon -
\frac{\bar{\Gamma}}{2} A(\omega)+i\frac{\bar{\Gamma}}{2}[1+f(\omega)]\right)^{-1}$, 
where
$ A(\omega)=\frac{1}{\pi}\left\{\psi\left(\frac{1}{2}+\frac{\beta E_{\text{c}}}
{2 \pi}\right)-\text{Re}\left[\psi\left(\frac{1}{2}+i \frac{\beta \omega}
{2 \pi}\right)\right]\right\}$,
and $\bar{\langle n\rangle}=-\int\frac{d\omega}{\pi}\text{Im}\left\{ \bar{G}_0^{\text{r}}(\omega)\right\}f(\omega)$ is the occupation of the level
per spin, $\psi$ the Digamma function,
and $E_\text{c}$ a high-energy cutoff. 
For the high-temperature limit, $\beta \bar{\Gamma} \ll 1$, we obtain the
analytical expression
\begin{equation}
\label{qhight}
   Q =- \frac{e \eta}{2} \left[ f''(\epsilon)+ 
    \frac{f'(\epsilon)}{1+f(\epsilon )} \left( f'(\epsilon)+
    \frac{2A(\epsilon)/\bar{\Gamma}}{1+f(\epsilon)}\right)\right] . 
\end{equation}

\begin{figure}
\includegraphics[width=3.in]{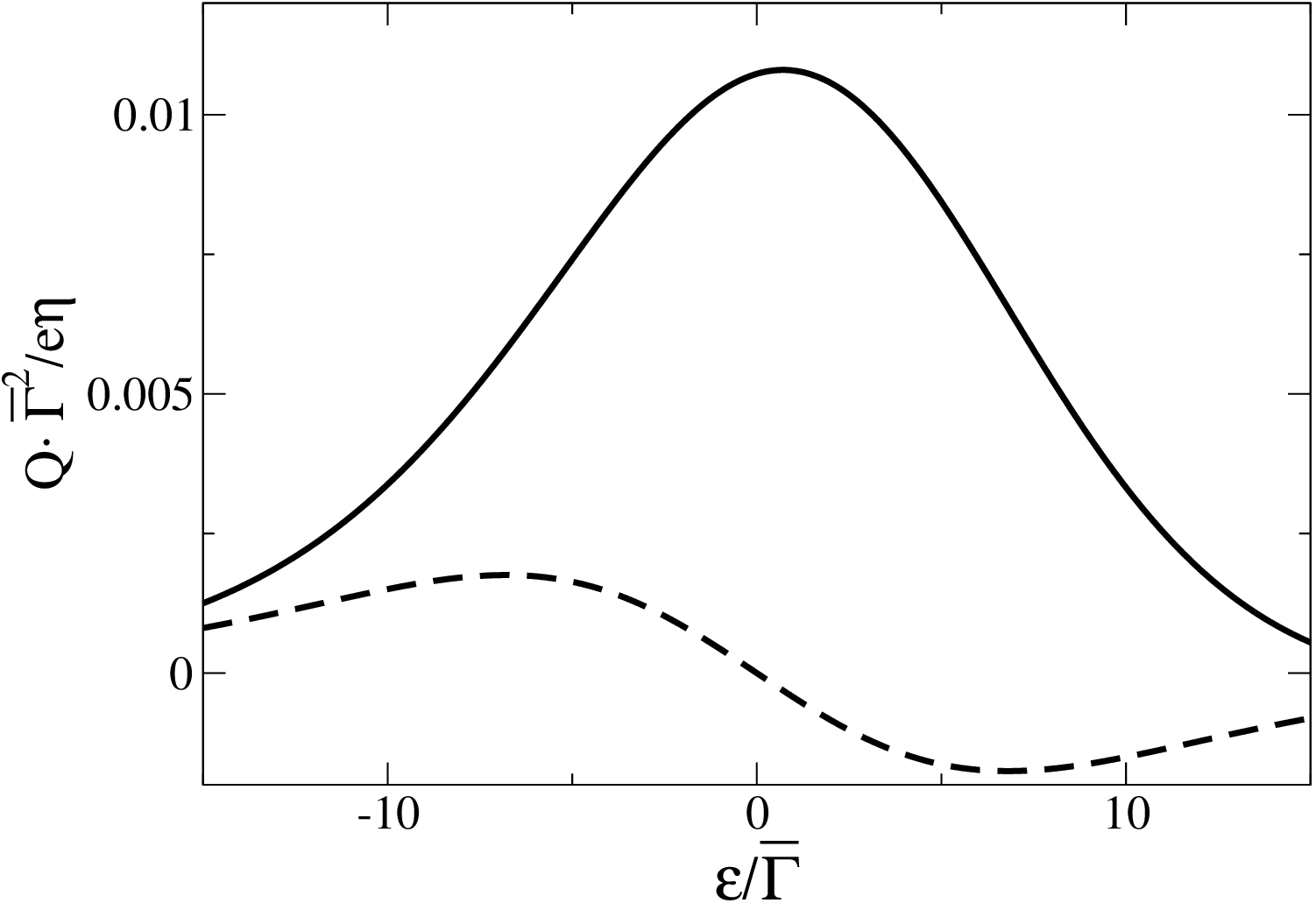}
\caption{
  Pumped charge in units of $e\eta/\bar{\Gamma}^2$ as a function of 
  the level position in units of $\bar{\Gamma}$ for $U\rightarrow \infty$ 
  (solid line) and $U=0$ (dashed line).
  The temperature is $k_{\text{B}} T=5\bar{\Gamma}$, and 
  $E_{\text{c}}=20\bar{\Gamma}$  .
\label{hightemp}}
\end{figure}

Figure~\ref{hightemp} shows the pumped charge as a function of the level 
position in the high-temperature limit. 
The enhancement of the pumped charge as compared to the 
noninteracting case,
is mainly due to the fact that, in the presence of interactions, 
the bare level is renormalized by an amount which depends on $\Gamma(t)$, and 
hence the level position becomes time dependent. 
The oscillation of the level increases the pump effect 
[third term in Eq.~(\ref{qhight})] \cite{note0}. 
Also the 
fact that the level width is energy dependent, $\frac{\bar{\Gamma}}{2} [1+f(\omega)]$, has some small effect on the pumped charge 
[second term in Eq.~(\ref{qhight})]. 
We note that for $\beta \rightarrow 0$, the third term in Eq.~(\ref{qhight}) 
goes as $\beta^2 \bar{\Gamma} E_{\text{c}}$, while the other two terms go as 
$(\beta \bar{\Gamma})^2$ \cite{note1}. 
The shape of the curves in Fig.~\ref{hightemp} are easily understood from the
dependence of $\partial \bar{\delta}/ \partial \bar{\Gamma}$ around $|\omega| \lesssim k_{\text{B}} T$  on
the bare level position $\epsilon$.
In the noninteracting case, the scale on which the phase
$\bar \delta$ varies around the level position $\epsilon$ increases linearly 
with $\bar\Gamma$; i.e., $\partial \bar{\delta}/ \partial \bar{\Gamma}$ changes sign when tuning the level position through the
Fermi energy. 
In presence of interaction, though, the dominant mechanism is the variation
of the level renormalization, which shifts $\bar \delta(\omega)$
along the $\omega$ axis, with no sign change in $\partial \bar{\delta}/ 
\partial \bar{\Gamma}$.

\begin{figure}
\includegraphics[width=3.in]{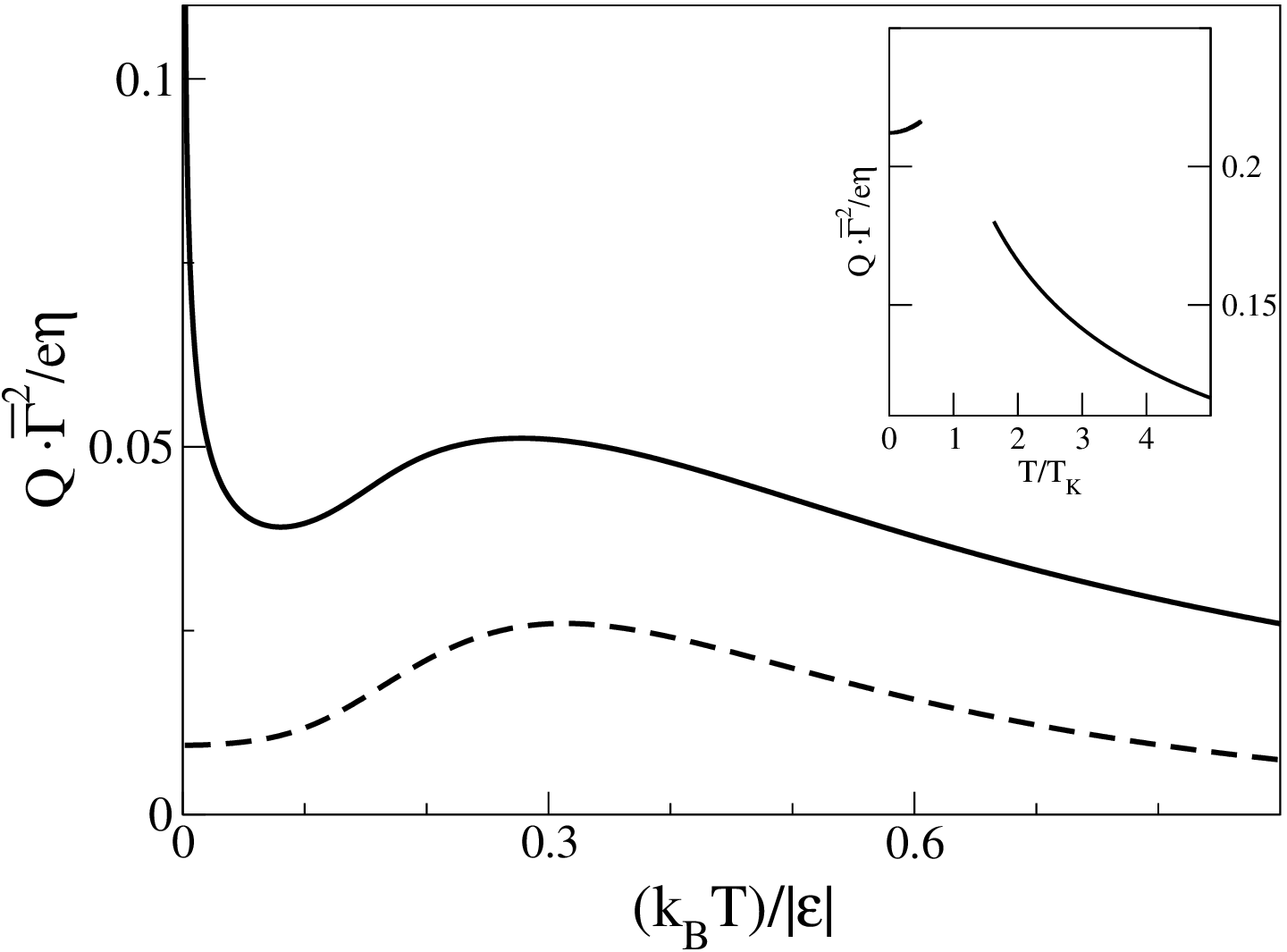}
\caption{
  Pumped charge in units of $e\eta/\bar{\Gamma}^2$ as a function of 
  temperature (in units of the level position), for $U\rightarrow \infty$ 
  (solid line) and for $U=0$ (dashed line). 
  The level position is fixed at $\epsilon=-2.5 \bar{\Gamma}$. 
  In both cases a maximum appears at almost the same temperature. 
The cutoff energy is  $E_\text{c}=20 \bar{\Gamma}$.\\
Inset: Pumped charge as a function of temperature 
(in units of 
$T_{\text{K}}$) for $U\rightarrow\infty$. 
For $0<T<0.5 T_{\text{K}}$, $Q$ is obtained by means of the mean-field 
slave-boson method
, performing a Sommerfeld expansion. 
For $T>1.5 T_{\text{K}}$ it is computed numerically 
using the equation-of-motion Green's function. 
\label{tempplot}}
\end{figure}

The equation-of-motion method gives qualitative, reliable results down to the 
Kondo temperature, given by
$k_\text{B}T\approx k_\text{B}T_{\text{K}}= \sqrt{E_{\text{c}}\bar{\Gamma}}/2
\exp\left( - \pi |\epsilon |/\bar{\Gamma}\right)$.
In Fig.~\ref{tempplot}, we show the temperature dependence of the pumped 
charge for the interacting quantum dot, obtained by numerical integration of
 Eq.~(\ref{weakpumping})
 and for comparison, the noninteracting result.
At very high temperatures the pumped charge tends to zero. 
Decreasing the temperature the charge exhibits a maximum both in the 
interacting and noninteracting case. It occurs when the level position and the temperature are of the same order.
Its position is determined by the spectral weight of the integrand function in 
Eq.~(\ref{weakpumping}) which falls in the energy window set by temperature 
through the derivative of the Fermi function. 
Approaching the Kondo temperature, the pumped charge in the interacting system 
increases rapidly, indicating Kondo correlations.\\
To address the limit $T\ll T_{K}$ we resort to the slave-boson method 
\cite{slaveboson} in the mean-field approximation in the boson field. 
The instantaneous dot Green's function can be written as 
$ G_0^{\text{r}}(\omega,t)=\frac{\Gamma_{\text{pf}}(t)}{\Gamma(t)} 
G_\text{pf}^{\text{r}}(\omega,t)$, where the pseudofermion Green's function is
$G_{\text{pf}}^{\text{r}}(\omega,t)=(\omega-\epsilon_{\text{pf}}(t)+i\Gamma_
{\text{pf}}(t)/2)^{-1}$.
We are interested in $\bar{G}_{\text{pf}}^{\text{r}}(\omega)$, where $\Gamma_{\alpha}(t)$ is replaced by $\bar{\Gamma}_{\alpha}$.
The renormalized level position $\bar{\epsilon}_{\text{pf}}$ and  
the renormalized rate $\bar{\Gamma}_{\text{pf}}$ have to be found as the solutions of  
the non-linear system of equations:  
\begin{subequations}
\label{selfconsistency}
\begin{eqnarray}
& &  2\bar{\Gamma} \int \frac{d \omega}{2 \pi} \text{Re}\left\{\bar{ 
G}_{\text{pf}}^{\text{r}}(\omega)\right\} f(\omega) + 
\bar{\epsilon}_{\text{pf}}-\epsilon=0,\\
& &  -4 \int \frac{d \omega}{2 \pi} \text{Im}\left\{ 
\bar{G}_{\text{pf}}^{\text{r}}(\omega)\right\} f(\omega)=1-\frac{\bar{\Gamma}_{\text{pf}}}{\bar{\Gamma}}.
\end{eqnarray}
\end{subequations}
At zero temperature the pumped charge can be expressed by means of Friedel's sum rule \cite{langreth} and Eq.~(\ref{weakpumping}), as  
\begin{equation}
Q=-\frac{4 e \eta} {\bar{\Gamma}} \frac{\partial\bar{\langle n\rangle}}{\partial 
\bar{\Gamma}} \sin^2\left(\pi \bar{\langle n\rangle}\right) \, ,
\end{equation}
which relates the pumped charge to the average occupation per spin 
$\bar{\langle n\rangle}$ only.
The full knowledge of the latter, e.g. from numerical renormalization group,
would establish an exact solution of the problem.
Within the mean-field slave-boson approach we get 
$\bar{\langle n\rangle}=1/2 (1-\bar{\Gamma}_{\text{pf}}/\bar{\Gamma})$ \cite{hewson}, and 
$\bar{\Gamma}_{\text{pf}}$ is computed from Eqs. (\ref{selfconsistency}).   
In the unitary limit ($-\epsilon \gg \bar{\Gamma}_{\text{pf}}$ and $T\ll T_\text{K}$,
such that $\bar{\langle n\rangle} \rightarrow 1/2$) the level is renormalized 
to resonance $\bar{\epsilon}_{\text{pf}}\rightarrow 0$, and the pumped charge is zero. 
This result is consistent with the fact that in the unitary limit the 
problem maps to the noninteracting dot with the level shifted to 
resonance, and that for the free-electron case there is no pumped charge 
when the level is at the Fermi energy. 
On the other hand, in experimentally relevant situations the renormalized level is not exactly at the Fermi energy and 
non-negligible charge pumping occurs.  
 
\begin{figure}
\includegraphics[width=3.in]{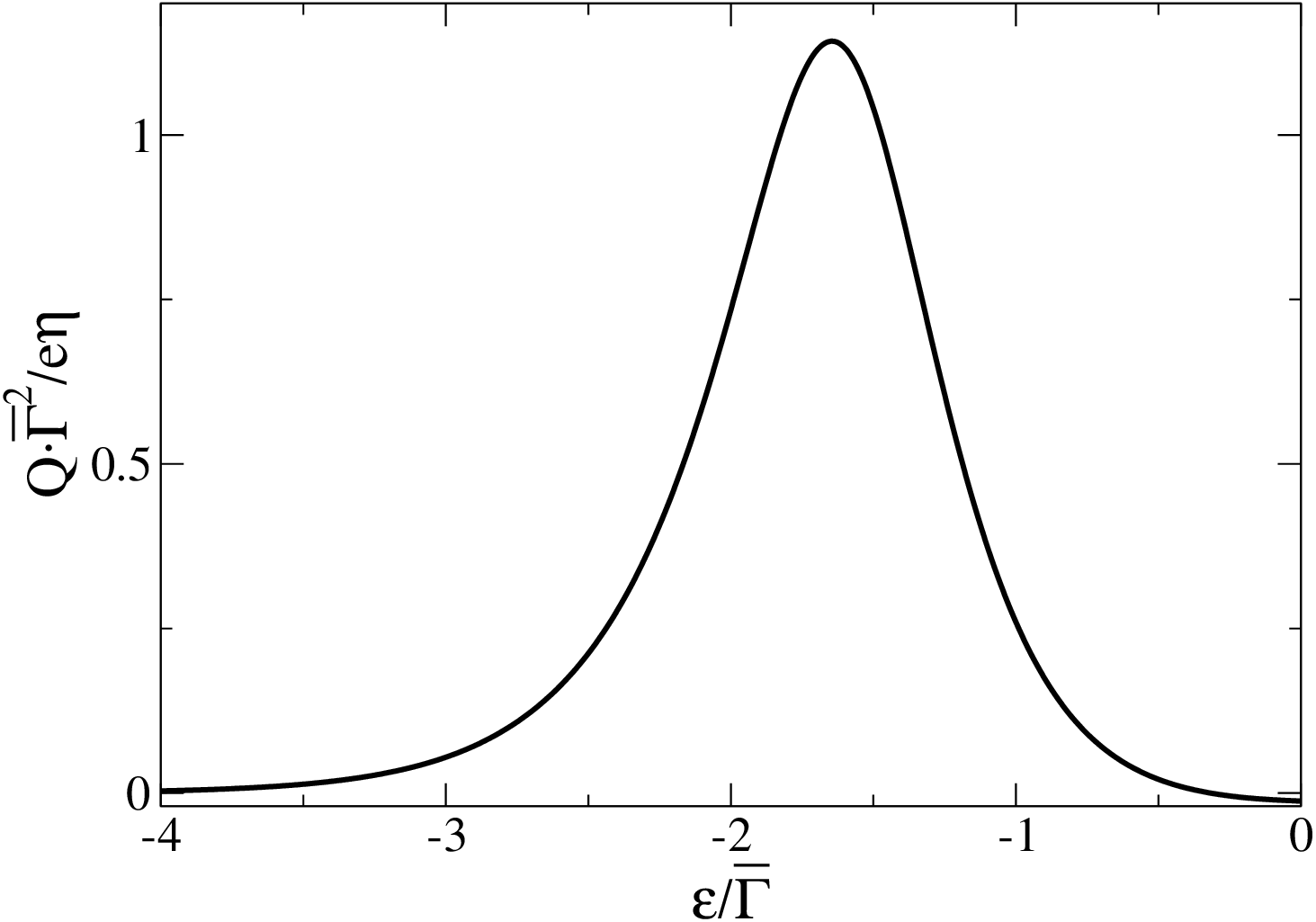}
\caption{
  Pumped charge in units of $e\eta/\bar{\Gamma}^2$ as a function of 
  the level position in units of $\bar{\Gamma}$, for $U\rightarrow \infty$, $T=0$ and $E_{\text{c}}=20 \bar{\Gamma}$,
  computed by means of the mean-field slave-boson method. 
  \label{zerotempfig}}
\end{figure}

In Fig.~\ref{zerotempfig} we show the charge pumped at zero temperature  
obtained solving numerically Eqs.~(\ref{selfconsistency}). 
As expected the charge 
tends to zero when the level is deep enough below the Fermi energy. 
The behavior of the pumped charge around $T\ll T_{\text{K}}$ can be obtained by
performing 
a Sommerfeld expansion in Eq.~(\ref{weakpumping}), and 
in Eqs.~(\ref{selfconsistency}). The pumped charge goes as $T^2$. 
Comparing the temperature behavior of the charge for $T\ll T_{\text{K}}$ with 
the one for $T$ just above $T_{\text{K}}$ [see inset of Fig.~\ref{tempplot}], 
we expect a maximum at around $T_{\text{K}}$. 
Roughly speaking, this extremum is analogous to the one that occurs at higher 
temperatures.

\textit{Acknowledgments.} 
We acknowledge useful discussions with E. Mucciolo, Y. Oreg, E. Sela, and F. Taddei, and   
support from DFG via SFB491 and GRK726 (J.K.) and from EC through grants EC-RTN Nano, EC-RTN Spintronics and EC-IST-SQUIBIT2 (M.G., J.S., and R.F.).


\begin{thebibliography}{24}

\bibitem{thouless} D. J. Thouless, Phys. Rev. B \textbf{27}, 6083 (1983). 

\bibitem{brouwer}
        P. W.~Brouwer, Phys. Rev. B {\bf 58}, R10135 (1998).

\bibitem{emissivity} M. B\"uttiker, H. Thomas, and A. Pr\^etre, Z. Phys. B \textbf{94}, 133 (1994).

\bibitem{zhou}
        F.~Zhou, B.~Spivak, and B.~Altshuler, Phys. Rev. Lett.
        {\bf 82}, 608 (1999); Yu. Makhlin and A. D.~Mirlin, Phys. Rev. Lett. {\bf 87},
        276803 (2001); O.~Entin-Wohlman, A.~Aharony, and
        Y. Levinson, Phys. Rev. B {\bf 65}, 195411 (2002); M.~Moskalets and M.~B\"uttiker, Phys. Rev. B {\bf 66},
        035306 (2002); {\bf 66} 205320 (2002). 


\bibitem{marcus}M. Switkes, C.M. Marcus, K. Campman, and A.C. Gossard, 
Science {\bf 283}, 1905 (1999); S.K. Watson, R.M. Potok, C.M. Marcus, and V. Umansky, Phys. Rev. Lett {\bf 91}, 258301 (2003).


\bibitem{pothier}
H. Pothier, P. Lafarge, C. Urbina, D. Esteve, and M.H. Devoret, Europhys. Lett. {\bf 17}, 249 (1992); I. L. Aleiner and A. V. Andreev, Phys. Rev. Lett. 
\textbf{81}, 1286 (1998); R. Citro, N. Andrei, and Q. Niu, Phys. Rev. B {\bf 68}, 165312 (2003); P. W. Brouwer, A. Lamacraft, and K. Flensberg, Phys. Rev. B {\bf 72}, 075316 (2005).


\bibitem{aono}
T. Aono, Phys. Rev. Lett. \textbf{93}, 116601 (2004).

\bibitem{cota}
E. Cota, R. Aguado, and G. Platero, Phys. Rev. Lett. {\bf 94}, 107202 (2005).




\bibitem{raikh}
L. I. Glazman and M. E. Raikh, JETP Lett. \textbf{47}, 452 (1988); T. K. Ng and P. A. Lee, Phys. Rev. Lett. \textbf{61}, 1768 (1988).


\bibitem{jauho}
A. P. Jauho, N. S. Wingreen, and Y. Meir, Phys. Rev. B \textbf{50}, 5528 (1994).

\bibitem{note_corrections}
The average-time approximation amounts in neglecting corrections to 
the self-energy due to the perturbation 
$[\tau- (t_1+t_2)/2]{\dot H}(t_0)$.
By retaining them one gets a correction to $ \check G_1(\omega,t)$, Eq.(\ref{g1}), 
which can be written as 
$\hbar \check G_0(\omega,t) \check\Sigma_{1}^{\text{corr}} (\omega,t) 
\check G_0(\omega,t)$. The subsequent correction to the current formula 
Eq.~(\ref{adiabaticcurrent}) depends on the vertex function [E. Sela and Y. Oreg, cond-mat/0509467 (2005)].
By comparing with a diagrammatic approach at high temperature [J. Splettstoesser \textit{et al}, in preparation] and relying on the mapping to noninteracting fermions at zero temperature, we 
can conclude that the neglected term leads to, at most, quantitative corrections.


\bibitem{note3} 
For the case that the dot is replaced by an arbitrary interacting region 
with many states, Eq.~(\ref{adiabaticcurrent}) is easily generalized by
replacing all Green's functions and $\Gamma(t)$ with matrices, performing
the trace, and dividing a factor of 2 if the spin index is accounted for 
in the matrix structure.

\bibitem{transmission}
Y. Meir and N. S. Wingreen, Phys. Rev. Lett. {\bf 68}, 2512 (1992);
J. K\"onig, H. Schoeller, and G. Sch\"on, Phys. Rev. Lett. {\bf 76}, 1715 
(1996);
J. K\"onig, J. Schmid, H. Schoeller, and G. Sch\"on, Phys. Rev. B {\bf 54}, 
16820 (1996).


\bibitem{eom}  Y. Meir,  N. S. Wingreen, and P. A. Lee,  Phys. Rev. Lett. 
\textbf{66}, 3048 (1991).


\bibitem{note0} It is known that for pumping through a noninteracting dot 
with one of the barriers and the level position, at high temperature, the 
pumped charge scales as $\beta \bar{\Gamma}$, in contrast to 
$(\beta \bar{\Gamma})^2$ for pumping with the two barriers. 

\bibitem{note1} 
A time-dependent level renormalization also occurs in the absence of 
interaction but with a non-constant density of state in the leads.
The extra contribution to the pumped charge, however, behaves as 
$\beta \bar{\Gamma}$ for $\beta \rightarrow 0$, and can, therefore, be
distinguished from the interacting behavior.


\bibitem{slaveboson} 
 P. Coleman, Phys. Rev. B \textbf{29}, 3035 (1984).

\bibitem{langreth}
D. C. Langreth, Phys. Rev. \textbf{150}, 516 (1966).

\bibitem{hewson}
A. C. Hewson, \textit{The Kondo Problem to Heavy Fermions} (Cambridge University Press, Cambridge England, 1993).
\end{thebibliography}
\end{document}